# The Potential of Exozodiacal Disks Observations with the WFIRST Coronagraph Instrument

*A white paper submitted in response to the National Academies of Science, Engineering and Medicine's Call on*
**Exoplanet Science Strategy**

Lead Author: Bertrand Mennesson, Jet Propulsion Laboratory, California Institute of Technology. Email: Bertrand.mennesson@jpl.nasa.gov
Phone: (818) 354-0494

Co-Authors: V. Bailey (JPL), J. Kasdin (Princeton), J. Trauger (JPL), R. Akeson (Caltech/IPAC), L. Armus (Caltech/IPAC), J. L. Baudino (Oxford/AOPP), P. Baudoz (Paris Obs./CNRS), A. Bellini (STScI), D. Bennett (NASA-GSFC), B. Berriman (Caltech/IPAC), A. Boccaletti (Paris Obs./CNRS), S. Calchi-Novati (Caltech/IPAC), K. Carpenter (NASA-GSFC), C. Chen (STScI/JHU), W. Danchi (NASA-GSFC), J. Debes (STScI), S. Ertel (Univ. of Arizona), M. Frerking (JPL), C. Gelino (Caltech/IPAC), D. Gelino (Caltech/IPAC), J. Girard (STScI), T. Groff (NASA-GSFC), S. Kane (UC Riverside), G. Helou (Caltech), J. Kalirai (STScI), Q. Kral (IoA Cambridge), J. Krist (JPL), J. Kruk (NASA-GSFC), Y. Hasegawa (JPL), A. M. Lagrange (IPAG/CNRS), S. Laine (Caltech/IPAC), M. Langlois (CRAL/CNRS), P. Lowrance (Caltech/IPAC), A. L. Maire (MPIA), S. Malhotra (NASA-GSFC), A. Mandell (NASA-GSFC), P. Marshall (NASA-GSFC), M. McElwain (NASA-GSFC), T. Meshkat (Caltech/IPAC), R. Millan-Gabet (GMT), L. Moustakas (JPL), B. Nemati (Univ. of Alabama), R. Paladini (Caltech/IPAC), M. Postman (STScI), L. Pueyo (STScI), E. Quintana (NASA-GSFC), S. Ramirez (Caltech/IPAC), J. Rhodes (JPL), A. J. E. Riggs (JPL), M. Rizzo (NASA-GSFC), D. Rouan (Paris Obs./CNRS), R. Soummer (STScI), K. Stapelfeldt (NASA-JPL), C. Stark (STScI), M. Turnbull (GSI), R. van der Marel (STScI), A. Vigan (LAM/CNRS), M. Ygouf (Caltech/IPAC), M. Wyatt (Univ. of Cambridge), F. Zhao (JPL), N. Zimmerman (NASA-GSFC)

## Acknowledgements
Part of this research was carried out at the Jet Propulsion Laboratory, California Institute of Technology, under a contract with the National Aeronautics and Space Administration.

**Summary**

The Wide Field Infrared Survey Telescope (WFIRST) Coronagraph Instrument (CGI) will be the first high-performance stellar coronagraph using active wavefront control for deep starlight suppression in space, providing unprecedented levels of contrast, spatial resolution, and sensitivity for astronomical observations in the optical. One science case enabled by the CGI will be taking images and (R~50) spectra of faint interplanetary dust structures present in the habitable zone of nearby sunlike stars (~10 pc) and within the snow-line of more distant ones (~20 pc), down to dust density levels commensurate with that of the solar system zodiacal cloud. Reaching contrast levels below ~$10^{-7}$ for the first time, CGI will cross an important threshold in debris disks physics, accessing disks with low enough optical depths that their structure is dominated by transport phenomena than collisions. Hence, CGI results will be crucial for determining how exozodiacal dust grains are produced and transported in low-density disks around mature stars. Additionally, CGI will be able to measure the brightness level and constrain the degree of asymmetry of exozodiacal clouds around individual nearby sunlike stars in the optical, at the ~10x solar zodiacal emission level. This information will be extremely valuable for optimizing the observational strategy of possible future exo-Earth direct imaging missions, especially those planning to operate at optical wavelengths, such as Habitable Exoplanet Imaging Mission (HabEx) and Large Ultraviolet/Optical/Infrared Surveyor (LUVOIR).

## 1. What is (exo)zodiacal dust and how does it impact exoplanet direct imaging?

The solar system zodiacal cloud contains a population of small (~1–100 μm) warm dust grains located within the asteroid belt, extending from <0.1 AU to ~3.3 AU. The COBE Diffuse Infrared Background Experiment (DIRBE) included measurements of the zodiacal light foreground from 1.25 to 240 μm, enabling modeling of the brightness distribution, grain size distribution, temperature, and optical depth radial profiles with high accuracy.[1,2] While its optical depth is only $10^{-7}$ at 1 AU and its total mass estimated to only a few $10^{-9}$ Earth mass—equivalent to an asteroid of 15 km in diameter—the zodiacal cloud integrated flux dominates that of any planet in the solar system at any wavelength ranging from the optical to the mid-infrared (IR).

Zodiacal dust is believed to originate in asteroid collisions and the evaporation/break-up of comets as they approach the Sun. While some authors suggest that most of the observed dust is of cometary origin, e.g., via spontaneous disruption of Jupiter family comets,[4] the relative contribution of asteroids and comets is still under debate. Similarly, "exozodiacal" dust refers to the inner (<few AU) warmer (>~200 K) part of circumstellar debris disks, where terrestrial planets form, and where we might see the signature of "exo-comets" and "exo-asteroids." Because zodiacal dust grain lifetimes are much shorter than stellar lifetimes, it is generally believed that exozodiacal dust must be regenerated (e.g., Ref. 5) to be observed around main sequence stars. The inner brightness distribution of equivalent

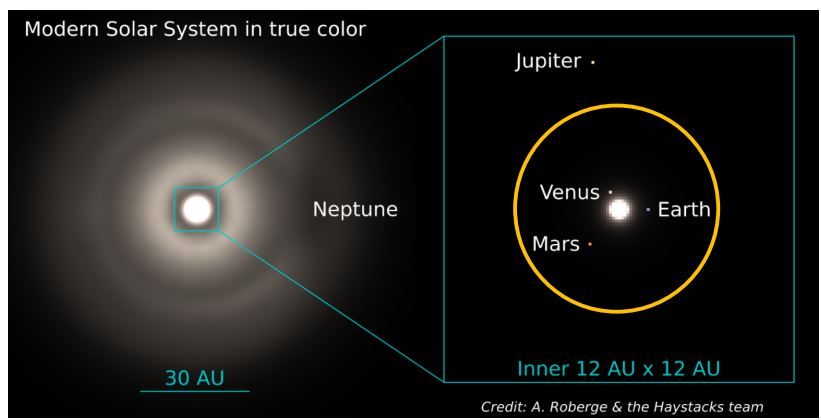

**Figure 1:** Model of the solar system as viewed pole-on in the visible with a spatial resolution of 0.03 AU per pixel and the Sun "removed".[3] *Left panel*: brightness scaling adjusted to highlight dust structures. *Right panel*: zoom-in of the inner region *with a steeper (non-linear)* brightness scaling revealing the Earth, Venus, and Mars. The solar zodiacal emission comes from regions within the asteroid belt, at the approximate distance represented by the yellow circle.

"exozodiacal" dust structures in debris disks around other mature stars is then expected to reflect present dust sources (comets, asteroids), as well as sinks (Poynting-Robertson drag, radiation pressure), and perturbations (collisions, evaporation, planets), revealing some of the system's current dynamical state and formation history. In particular, bright exozodiacal disks may be the signposts of outer planets scattering numerous comets in the inner regions, similar to what happened during the solar system Late Heavy Bombardment (e.g., Refs. 4, 6).

But the presence of exozodiacal dust is really a double-edged sword. Indeed, bright exozodiacal dust structures can provide key information about the dynamical processes at play in other planetary systems, but at the same time, they may represent a significant impediment to the direct imaging and spectral characterization of planets around other stars, particularly any faint Earth-like exoplanets orbiting in their habitable zone (HZ). Considering for instance a 4 m telescope viewing a Sun-Earth twin system at 10 pc with an exact replica of the solar zodiacal cloud, the corresponding exozodiacal dust flux *per spatial resolution element* (PSF FWHM) is a few-hundred times brighter than the Earth at 10 μm (e.g., Refs. 7, 8), and still ~3 times brighter than the Earth seen at quadrature in the visible.[9] A bright exozodiacal disk will contribute a higher background noise and increase the exposure time required for exoplanet direct detection. Realistic and optimized observing scenarios for exo-Earth direct-imaging missions[10] estimate that a factor of 10 increase in exozodiacal dust density level, e.g., from solar level (1 "zodi") to 10 times higher (10 "zodis"), reduces the exo-Earth yield of such missions by a factor of ~2. While manageable, this loss in sensitivity is still significant. A potentially more problematic effect of bright exozodiacal emission is the creation of bright "clumps," regions of density enhancement trailing and leading the planet in its orbit, as predicted by disk-planet interaction models and actually observed in the solar system.[1] Simulations conducted in the case of an Earth analog embedded in exozodiacal clouds of different brightnesses[11] predict for instance that at a level of 20 "zodis," local heterogeneities in the disk could be brighter than an exo-Earth and constitute important sources of confusion and false positives. The exact location and strength of these clumps is expected to vary with planet mass, semi-major axis and outer dust characteristics, e.g., density and typical grain size.[12] However, the main result is that exozodiacal clouds at density levels of ~20 solar zodis or more may generate bright enough clumps to preclude the detection of exo-Earths, or at least make data interpretation difficult, especially for systems seen at high inclination. In highly inclined systems, we must look through a much larger column of dust to see planets, including any cold dust in the system. Although one may think this cold dust would be negligibly faint due to the $1/r^2$ illumination factor, forward scattering of light by dust grains can partially counter this effect. As a result, the surface brightness in an edge-on HZ can be dominated by dust physically located beyond a few AU instead of dust within the HZ.[13] A better knowledge of exozodiacal disk brightness level and morphology—both per individual star and in a statistical sense—plays an important role for optimizing future space missions aiming at the characterization of Earth-like exoplanets, as recognized early on (e.g., Refs. 11, 14).

## 2. Current state of knowledge and limitations of existing facilities

If exozodical clouds are similar to what is observed in the solar system, their spatially integrated flux—relative to the central star—is roughly 1,000 times brighter in the mid-IR than in the visible. For a perfect solar system analog seen pole-on for instance, the ratio of total zodiacal flux to stellar flux is ~$4\times10^{-5}$ at 10 μm, compared to only ~$4\times10^{-8}$ at V band. This advantage in contrast is well known, and *in the absence of a high-contrast visible space-based coronagraph*, exozodiacal surveys have been primarily conducted in the *mid-IR*, either through space-based spectroscopic measurements, or through ground-based spatially resolved measurements.

Space-based infrared telescopes, such as the Infrared Astronomical Satellite (IRAS), the Infrared Space Observatory (ISO), and Spitzer, are too small to spatially separate the exozodiacal dust emitting region from the central star. They rely instead on spectral excess measurements, which require careful calibration and accurate subtraction of the model dependent stellar spectral energy distribution. As a result, Spitzer detection limits for exozodiacal disks are typically 100 zodis at 24 μm, and 1,000 zodis at 10 μm, the wavelength most sensitive to HZ dust. Only a few warm excesses have been detected by Spitzer around mature stars above these detection limits.[15,16] Out of 203 FG main sequence stars observed with the Spitzer InfraRed Spectrometer (IRS), only two showed an excess in the short wavelength band (8.5–12 μm).[17]

To detect and characterize exozodiacal disks around a large number of sunlike stars *in the mid-IR*, significantly lower dust density levels must be accessed, and the dust-emitting region needs to be spatially resolved from the star. This calls for improvements in both contrast and spatial resolution and means that nulling interferometry is required. Three instruments have tackled this observational challenge over the last 20 years: the Multi Mirror Telescope Nuller,[18,19] the Keck Interferometer (KI) Nuller,[20,21] and the Large Binocular Telescope Interferometer (LBTI).[22,23] The KI observations reached a typical detection limit of ~300 to 500 zodis per star[24,25] between 8 and 10 μm. The LBTI exozodi key science survey, which started in 2013,[26,27] has been demonstrating further improvements in sensitivity. The statistical analysis of LBTI data obtained to date indicates with 95% confidence that the typical (median) level of exozodi emission around sunlike stars with no outer cold dust reservoir previously known is below 26 zodis.[28] This number, which benefits from averaging over a few dozen stars, should not be confused with the detection limit *per individual star*, which is about 30 zodis for early spectral types and ~100 zodis for solar analogs.

While the LBTI sensitivity is within a factor of 5-10 of the confusion limit proposed by Defrere et al.,[11] there are limitations of this ground-based mid-IR approach including bright thermal (sky) backgrounds and insufficient spatial resolution to resolve disk substructures. More importantly, basic dust properties (e.g., density profile and size distribution) cannot be uniquely derived from measurements over a narrow wavelength range. This means that the brightness of exozodiacal emission at visible or shorter IR wavelengths cannot be reliably extrapolated from mid-IR measurements. This last issue is clearly illustrated by the intriguing detection of ~1% near-IR excesses around ~20% of main sequence stars,[29,30,31,32] with generally no detection counterpart in the mid-IR,[25] pointing to populations of very hot and small (submicron) grains piling up very close to the sublimation radius around these stars (e.g., Refs. 33, 34, 35). Visible exozodi observations are required[36] to measure the scattering phase function of dust grains, to better inform grain size and shape, and finally to enable connecting scattered light spectra with IR spectral energy distribution (SED) for compositional modeling.

In short, to make further progress, high-contrast, high spatial resolution space-based observations are required, ideally at multiple wavelengths. The WFIRST CGI will start this journey toward exozodiacal dust imaging and spectral characterization at low dust density levels (<~100 times solar), crossing an important threshold in debris disk physics, at a spatial resolution improved to ~50 mas in the visible.

## 3. How will CGI improve the state of knowledge and impact future missions?

The benefits of using coronagraphy for the study of exoplanetary systems became obvious with the first optical images of beta Pictoris's extended edge-on circumstellar disk obtained by Smith & Terrile.[37] Following the IRAS satellite discovery of a large IR excess around this star, these optical coronagraphic observations provided the first direct confirmation of planet formation and resolved images of dusty debris disks in another system. With the access to space provided by the Hubble Space Telescope (HST), many more (bright) circumstellar disks have been spatially resolved since (e.g., Refs. 38, 39, 40, 41). However, HST's high contrast instruments have only achieved high contrast at large separations, such as the $10^{-9}$ contrast detection of Fomalhaut b at 12".[42] The Space Telescope Imaging Spectrograph (STIS) is the only remaining operational high dynamic range optical instrument in space today. HST/STIS does provide access to separations as small as 0.25" in the visible, but only at ~$10^{-4}$ contrast,[43] limiting exozodiacal observations to the closest stars and again to disks with very high surface brightness, typically $10^4$ higher than in the solar system.

Table 1: List of currently envisioned WFIRST/CGI science filters and observing modes, together with contrast performance best estimates (CBEs) for each. Point source detection limits indicate the flux (relative to the central star) of the dimmest point source that can be detected at 5σ or higher *anywhere* within the range of angular separations indicated. Only the 3 highlighted modes will be *fully* tested from the ground, but all filters and modes will be available for CGI observations.

| CGI Filters | $\lambda_{Center}$ (nm) | BW | Channel | Masks | Working Angle (λ/D) | Working Angle (") | Point Source Detection Limit (CBEs) | Starlight Suppression Region |
|---|---|---|---|---|---|---|---|---|
| 1 | 575 | 10% | Imager | HLC | 3–9 λ/D | 0.15–0.43" | $1.5×10^{-9}$ | 360° |
| 2 | 660 | 18% | IFS | SPC | 3–9 λ/D | | | 130° |
| 2 | 660 | 18% | Imager | SPC | 3–9 λ/D | | | 130° |
| 3 | 760 | 18% | IFS | SPC | 3–9 λ/D | 0.20–0.56" | $3.5×10^{-9}$ | 130° |
| 3 | 760 | 18% | Imager | SPC | 3–9 λ/D | | | 130° |
| 4 | 825 | 10% | Imager | HLC | 3–9 λ/D | | | 360° |
| 4 | 825 | 10% | IFS | HLC | 3–9 λ/D | | | 360° |
| 4 | 825 | 10% | IFS | SPC disk | 6.5–12 λ/D | | | 360° |
| 4 | 825 | 10% | Imager | SPC disk | 6.5–12 λ/D | 0.50–1.46" | $6×10^{-10}$ | 360° |

With expected (current best estimates) point source detection limits better than $1.5×10^{-9}$ for angular separations from 0.15" to 1.46" at wavelengths ranging from 575 nm to 825 nm (Table 1), the WFIRST CGI promises a drastic improvement in high-contrast astronomical imaging capabilities at optical wavelengths. At 575 nm for instance, a point source detection limit of $1.5×10^{-9}$ is expected to be reached as close as 150 mas from the star, a separation referred to hereafter as the coronagraph inner working angle (IWA). In comparison, radiometric calculations (e.g., Ref. 44) indicate that for a solar system zodiacal cloud analog seen around a sunlike star at 7 pc at 575 nm and viewed under a 60 deg inclination with a 2.4m telescope, the disk flux contributed per spatial resolution element at the IWA (~1 AU in that case) is about $5×10^{-10}$ relative to the star. In other words, for a sunlike star at 7 pc, the detection limit is about 3 zodis at 1 AU, and would degrade to ~12 zodis at 2 AU, assuming the same instrument contrast and a quadratic fall-off in dust density. For sunlike stars close enough that some of their exozodiacal light emission can be captured at the CGI 150 mas IWA (i.e., located within ~20 pc), the dust surface brightness at the physical IWA will decrease as $1/d^2$. But so does stellar flux. As a result, for sunlike stars within ~20 pc, the sensitivity will be a constant 3 zodis at the IWA, reached at a physical separation of 1AU*(d/7pc). For cooler (respectively hotter) stars than the Sun,[45] the exozodi surface brightness at the Earth equivalent insulation distance would be lower (respectively higher), and the detection limits in zodi units would be slightly worse (respectively better). Figure 2 shows an illustration of the exozodiacal dust disk image expected when observing a nearby sunlike star with only 10 zodis of dust with the CGI hybrid Lyot coronagraph (HLC) mask at 575 nm. Dust emission at that level is clearly detected, together with some of the ring structures expected to be created by in-spiraling dust trapped in resonance with the orbit of a hypothetical perturbing planet.

## Conclusion

The simulations in Figure 2, which assume the current best estimates of coronagraph performance indicated in Table 1, illustrate the power of conducting sensitive spatially resolved exozodiacal observations with the WFIRST CGI, searching at the same time for the presence of otherwise undetectable planets (early type stars), possibly constraining their mass and orbit via their resonant structures, or the clearing of the inner disk. CGI will offer the first opportunity to explore disk-planet interactions at low dust density levels (~10× solar), and at very small physical separations: within the snow-line of sunlike stars located within ~20 pc, and in the HZ of those closer than ~10 pc.

The CGI will also directly measure exozodi levels and structures at the visible wavelengths considered for future missions for the first time. These observations will start to establish whether the zodiacal cloud of our inner solar system is representative of the population of our nearest sunlike neighbors; this information cannot be obtained from the ground. The CGI will hence pave the way to even more capable future direct imaging missions, as illustrated by the ongoing Starshade Probe, HabEx, and LUVOIR concept studies. With higher throughput, spatial resolution and contrast, such missions could thoroughly investigate the linkage between individual planet properties, planetary system architectures, and dust structures.

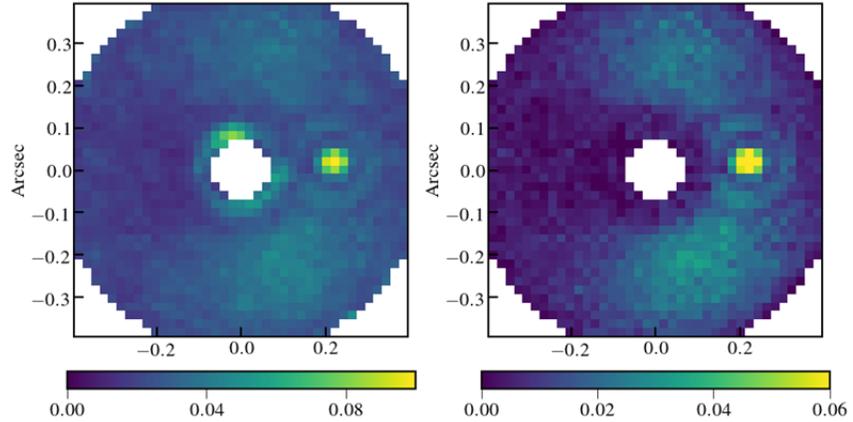

**Figure 2:** Simulated WFIRST CGI observations (HLC 575 nm imaging mask) of a nearby sunlike star (1 Ori, spectral type F6V at ~8 pc) hosting an exozodi dust cloud 10× denser than in the solar system, showing resonant structures due to a hypothetical jovian planet located at 1.6 AU. Flux scale is square-root stretch in units of photoelectrons/s. Simulated exposure time is 2.8 h. (Courtesy of M. Rizzo, N. Zimmerman and the "Haystacks" team). The right image shows the contrast enhancement provided by PSF subtraction (speckles removal) using observations of a reference star. The field of view diameter is 0.8" in both images.